\documentclass[conference]{IEEEtran}
\IEEEoverridecommandlockouts
\usepackage{cite}
\usepackage{amsmath,amssymb,amsfonts}
\usepackage{algorithmic}
\usepackage{graphicx}
\usepackage{textcomp}
\usepackage{xcolor}
\def\BibTeX{{\rm B\kern-.05em{\sc i\kern-.025em b}\kern-.08em
    T\kern-.1667em\lower.7ex\hbox{E}\kern-.125emX}}
\begin{document}

\title{Network Coding in Photonic-land: Three Commandments for Future-proof Optical Core Networks \\
}

\author{\IEEEauthorblockN{Dao Thanh Hai}
\IEEEauthorblockA{\textit{Posts and Telecommunications Institute of Technology} \\
Hanoi, Vietnam \\
haidt102@gmail.com}
%\and
%\and
%\and
%\and
%\and
}

\maketitle

\begin{abstract}
The digital transformation has been underway, creating digital shadows of (almost) all physical entities and moving them to the Internet. The era of Internet of Everything has therefore started to come into play, giving rise to unprecedented traffic growths. In this context, optical core networks forming the backbone of Internet infrastructure have been under critical issues of reaching the capacity limit of conventional fibers, a phenomenon widely referred as capacity crunch. For many years, the many-fold increases in fiber capacity is thanks to exploiting physical dimensions for multiplexing optical signals such as wavelength, polarization, time and lately space-division multiplexing using multi-core and/or few-mode fibers and such route seems to come to an end as almost all known ways have been investigated. This necessitates for a departure from traditional approaches to use the fiber capacity more efficiently and thereby improve economics of scale. Indeed, recent leaps-and-bounds progresses in photonic signal processing is expected to pave the way for re-defining the optical networking realm by transforming the convention functions of intermediate nodes from simply regeneration and/or switching to incorporating photonic signal processing functions, notably encoding and decoding ones. To this end, this paper lays out a new perspective to integrate network coding (NC) functions into optical networks to achieve greater capacity efficiency in a pragmatic manner by upgrading intermediate nodes functionalities. In addition to the review of our recent proposals on creating new research problems enabled by NC operations in optical networks, we also report state-of-the-art findings in the literature in an effort to renew the interest of network coding in optical networks and discuss three critical points for pushing forward its applicability and practicality including \textit{i)} network coding as a new dimension for multiplexing optical signals \textit{ii)} algorithmic aspects of network coding-enabled optical networks design \textit{iii)} network coding as an entirely fresh way for securing optical signals at physical layers. 
\end{abstract}


\section{Introduction}

The Internet of the future will be evolving to keep pace with the proliferation of digital innovations in all shapes and sizes \cite{ir4}. On one hand, from the user sides in broadest senses, there has been accelerated development of Internet usage and bandwidth-intensive services including such as autonomous vehicles, remote robotic surgery, tele-presence, to name but a few and of course the future unknown ones. A case for illustration is the operation of an autonomous car will generate about 5 Terabytes of data and it is no less than a current supercomputer in terms of generating and transmitting a colossal amount of data. On the other hand, thanks to the convergence of technologies, the massive adoption for new services will take more and more shorter time and it means that what is rare or unimaginable today can become widely adopted tomorrow. Such trend can be observed from the entire migration of working, learning and several other conventional physical activities to the Internet platform during the COVID-19. In this context, Internet traffic have been exploding and will continue to rise dramatically \cite{hai_oft2, hai_optik}. \\

Behind the scenes to support such exponential Internet traffic growth is the fiber-optics networks. Indeed, fiber-optic communications make up the backbone of Internet infrastructure, providing ultra-high capacity, low-latency and highly secured communication channels and hence, enabling the coming into availability of digital society and digital future \cite{ir4, hai_thesis}. For many years since its birth in 1966, fiber-optical communications have undergone remarkable progresses in capacity, spectrum efficiency and ultimately cost to keep pace with increasing demand in traffic growth and this is owning to many ground-breaking scientific and technological advances in the field of both electronics, photonics and digital signal processing \cite{hai_iet, hai_wiley, hai_ps1, hai_ps2, hai_csndsp, hai_icist1, hai_icist2, hai_sigtel1, hai_sigtel2, hai_icact, hai_nics, hai2021development}. In a span of roughly 30 years from 1990 to 2020, a 400 times increase in data rate has been recorded for a typical single wavelength channel, a spectacular rise from 2.5 Gb/s in around 1990 to 1 Tb/s in 2020 \cite{20years}. In addition to increase in data rate per channel, the number of channels per fiber is also multiplied via various multiplexing techniques to further improve overall transmission capacity. In particular, four physical dimensions of a light wave traveling along a fiber has been exploited including amplitude and phase, polarization, wavelength and lately space-division multiplexing by using multiple single-mode fibers in a same fiber cable or multi-core/few-modes fibers \cite{xor3}. Continuing the direction of improving multiplexing techniques and/or increase bit-rate per channel then appears to be increasingly difficult as hard limitation has been almost reached \cite{crunch}. On the constraint of finite spectrum capacity, innovative techniques and non-conventional ideas are needed to break from the norm in achieve higher capacity at greater economic of scales.  \\

Network coding (NC), originally invented in \cite{NC}, has soon become a revolutionary technique in networking to attain greater throughput, security and capacity. The core idea is that intermediate nodes, instead of simply storing and broadcasting data as in conventional networking paradigm, is allowed to manipulate data and then forward such (non-) linearly combined data to its output. The successes of network coding has been remarkable and therefore has been a \textit{de facto} in future wireless networks. However, with radically different transmission conditions, the wisdom from wireless networks could not be directly applied to optical networking realms. In the past due to the immature of photonic signal processing technologies, the proposal of exploiting NC in optical networks remains inadequately addressed due to the limited capability if network coding is performed in electronic domain. Nevertheless, the considerable advances in photonic signal processing and enabling technologies animates the interest in leveraging photonic network coding to re-define optical networking realm towards greater cost and energy-efficiency. Indeed, recent works in \cite{hai_comletter, hai_access, hai_oft, hai_comcom, hai_comcom2, hai_springer, hai_rtuwo, hai_systems, hai_springer2, hai2021consolation, nc_others9, nc_others7, icc} have renewed the interests and potential benefits of adding network coding layer to optical networks. The use of digital all-optical physical-layer network coding has also been extended to mm-wave radio-over-fiber networks \cite{nc_others1}, in visible light communications \cite{nc_others2, nc_others3}, and in passive optical networks (PON) \cite{nc_others4}. Data-center networking has been an active area for applying NC to reduce traffic \cite{nc_others5, nc_others6, nc_others8}. Physical layer encryption with NC has been also addressed in \cite{nc_security1, nc_security2}.  \\

In this paper, we argue that Network Coding and its implementation in photonic domain thanks to the maturing of enabling technologies could be a new venue to push forward the capacity boundary of a conventional optical fiber. We lay out a framework via an illustrative example to highlight potential uses cases of NC in optical networks relying on the core idea of transforming the functionalities of intermediate nodes from traditional tasks of regeneration and/or switching to encoding/decoding. The final point we raise is about the algorithmic aspects when it comes to network design and planning with network coding-enabled realm to reach its full potential gains. In supporting our arguments, we include some recent research results highlighting the impact of re-designing optical transport networks with NC compared to the conventional ones. \\
\section{Network Coding-enabled Optical Transport Networks: A New Paradigm}

Let us consider an exemplary case highlighting how network coding could be spectrally beneficial to optical networking. Assuming that there are two demands with dedicated protection from node $A$ and node $B$ to node $C$ as shown in Fig. 1. Their working signals and protection signals are indicated as $A_w$, $B_w$ and $A_p$, $B_p$ respectively. Supposing that the illustrative optical networks in Fig. 1 adopting the opaque architecture and thus, the protection signals from node $A$ and $B$ being routed via node $X$ are undergone optical-electrical-optical conversion before forwarding on fiber link $XC$. In the conventional operation, the elements composing of node $X$ is illustrated in Fig. 2. It can be seen that two optical transponders are needed to handle the protection signal $A_p$ ($B_p$) when it is routed over node X. From the functionality perspective, node $X$ simply performs regeneration functions on individual input signals and forward such regenerated signals to proper output links. We now turn attention to the new scenario where the intermediate node (node $X$) is powered by the new capability of mixing/encoding input signals and this is highlighted in Fig. 3. In this scenario, instead of simply processing input signal individually, the intermediate node $X$ performs XOR-encoding between protection signal $A_p$ and $B_p$ to create an encoded one $A_p \oplus B_p$. Such encoded $A_p \oplus B_p$ signal is then modulated by simply using one transponder instead of two transponders as in traditional case and is forwarded on link $XC$. Furthermore, on fiber link $XC$, rather than using two wavelength units to support $A_p$ and $B_p$, only one wavelength unit is needed to accommodate the encoded $A_p \oplus B_p$. At the destination node $C$, three signals are received including $A_w$, $B_w$ and $A_p \oplus B_p$ and it is important to note that in case of any single link failures, the receiver is fully capable of recovering the lost signal from the two remaining ones. The decoding process is shown in Fig. 4. \\

\begin{figure}[!ht]
	\centering
	\includegraphics[width=\linewidth]{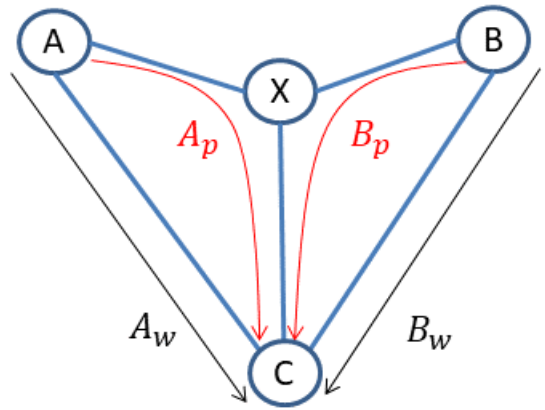}
	\caption{Two Traffic Demands with Dedicated Protection}
	\label{fig:i1}
\end{figure}

\begin{figure}[!ht]
	\centering
	\includegraphics[width=\linewidth, height = 6.5cm]{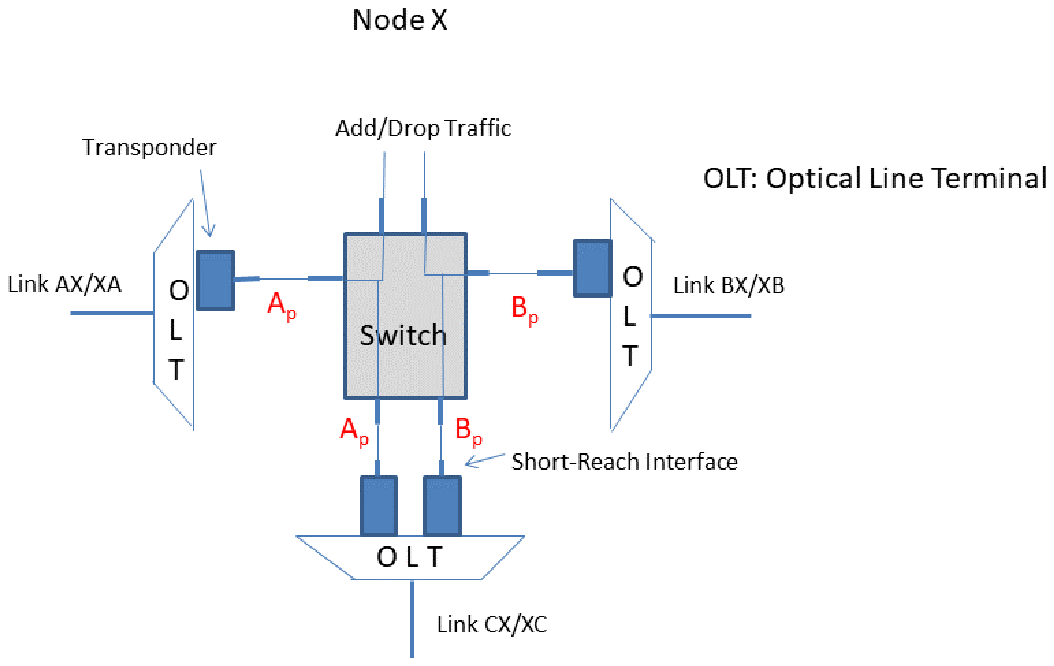}
	\caption{Optical Node X in Opaque Architecture Without Network Coding}
	\label{fig:i2}
\end{figure}

\begin{figure}[!ht]
	\centering
	\includegraphics[width=\linewidth, height = 6.5cm]{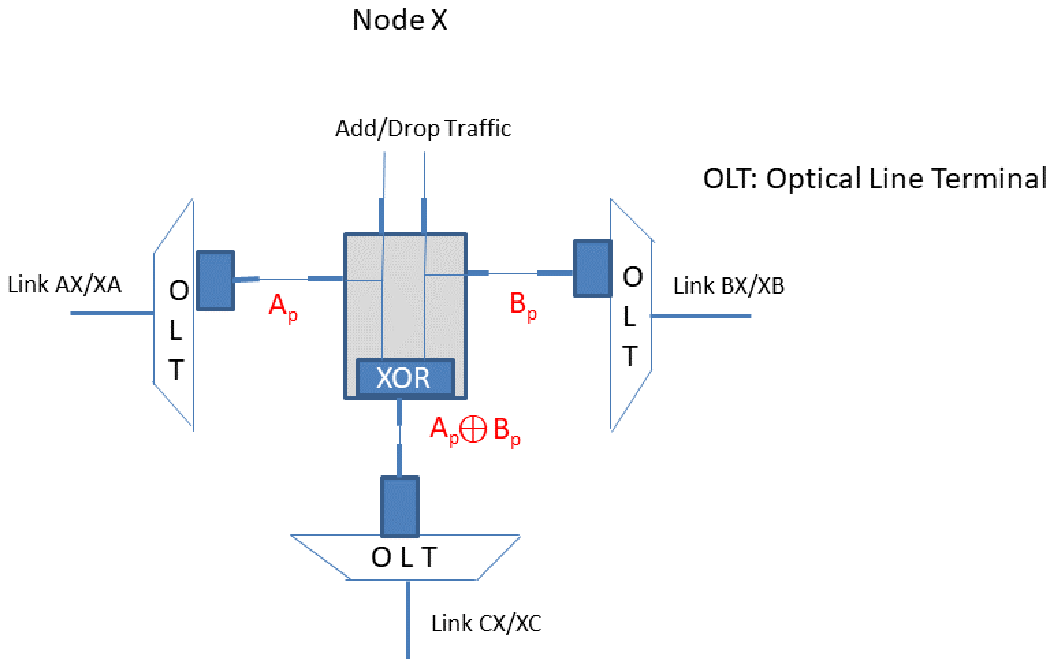}
	\caption{Optical Node X in Opaque Architecture With Network Coding}
	\label{fig:i3}
\end{figure}

\begin{figure}[!ht]
	\centering
	\includegraphics[width=0.7\linewidth, height = 5cm]{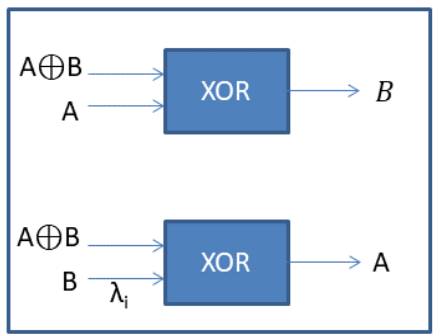}
	\caption{Decoding at Receiving Node upon Failure Events}
	\label{fig:i4}
\end{figure}

What the aforementioned example tells us is that network coding opens up new opportunities to re-define the optical networking realm radically from design and planning to operation and management. Among potential benefits, as clearly revealed in the example, the capital cost associated with transponders and spectrum usage could be remarkably reduced. Furthermore, as a consequence of hardware and spectrum savings, operational expenditure could also be reduced by turning off un-used hardwares including such as transponders and/or in-line amplifiers. The first commandment for applying network coding could be therefore stated that photonic network coding should be viewed as a new venue for multiplexing optical signals, in addition to well-known dimensions including wavelength, time, polarization and space and exploiting this new dimension could potentially heralds future-proof core networks that can sustain explosive traffic growths. \\

With respect to the use of XOR coding, it has to be noted that there has been remarkable progresses in photonic XOR which permits fully optical encoding/decoding up to Terabit/s \cite{nc_others10} and moreover, such operations could be optically performed between different modulation format signals \cite{nc_others11}. For the possible concern on time overhead of encoding and decoding processes, the fact that the encoding and decoding are all performed in the photonic domain constitutes thus negligible impact to the recovery time. This is radically different with other network coding schemes proposed in the literature that is performed in electronic domain in conjunction with the opaque network architecture. \\

%Its operations are typically based on semiconductor optical amplifiers (SoA) \cite{xor3} with advantages of low-power consumptions, easy deployment and short latency. Note that the state-of-the-art photonic XOR operation could be performed at the very high line rate up to 100 Gbps and with different modulation schemes such as BPSK and QPSK \cite{xor2}, \cite{xor3} 

\begin{figure}[!ht]
	\centering
	\includegraphics[width=0.7\linewidth, height = 6cm]{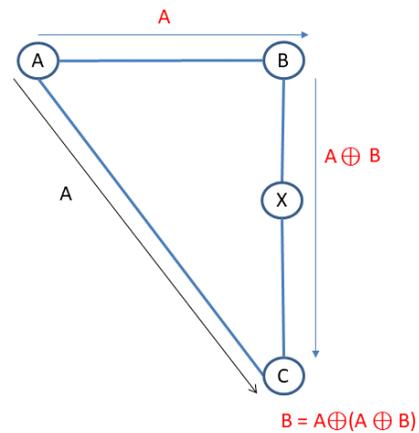}
	\caption{Physical Layer Security At Scale with Network Coding}
	\label{fig:i5}
\end{figure}

Next, let's move to a scenario where the introducing of network coding functions at optical nodes could drive the coming into availability of new services. Fig. 5 illustrates the case where there are two traffic demands from node $A$ and node $B$ to node $C$ respectively. However, different from traditional requests, demand $B$ is required an added service that the signal transmitted from node $B$ to node $C$ must be protected against physical eavesdropping and/or attacking. In offering such services, the optical signal on the route from node $B$ to node $C$ must be physically encrypted so that the tapping of such signal at intermediate nodes or in the middle way could not extract meaningful information. In this context, photonic network coding could be leveraged at node $B$ to create a physically encrypted signal, that is, $A \oplus B$ and transmit such encrypted one all the way to the receiving node $C$. By transmitting an encrypted version, the traffic from node $B$ to node $C$ could not be eavesdropped or decoded by any means. Only at the receiving node, the original signal is decoded with the key from the complementary signal, that is, $B = A\oplus(A\oplus B)$. This example highlights a radically original approach for providing security services at the physical layer without overhead in transmission as still only one wavelength unit is needed. Moreover, it has to be noted that such secured transmission is performed at a scale of a wavelength capacity which could be potentially up to Tb/s. In acknowledging this unique opportunity, the second commandment for application of photonic network coding is that there is a tremendous opportunity in upgrading the security of optical networks in a cost and energy-efficient manner and that could be performed at a scale of wavelength granularity. \\

It has become clear that the incorporation of photonic network coding paves the way for re-defining the optical networks to achieve greater efficiency in terms of capital, operational and security aspects. However, it should be noted that in order to realize such benefits, in addition to the technological upgrades, the algorithmic factors are of crucial importance. It has been well-known that algorithms involving all phases of a network from designing, planning to operation and management. Clearly a good set of algorithms result in better network utilization, even optimal one which in turn translates to considerable gain in capital and operation revenues. Conversely, poor algorithms could under-utilize the resources and as a consequence, potential benefits are failed to realized \cite{hai2021consolation, hai2021shades, hai2021achilles, hai_soict, hai2021pragmatic}. Well-designed algorithms are therefore critical to optimize the resources in optical networks. Although network coding could be a potentially new venue for leveraging optical networks, it indeed adds a new layer of complexity when it comes to network designs and operations. The introduction of network coding functions in optical networks gives rise to the issue of network coding assignment in which the determination of pair of demands for encoding/decoding, the coding nodes and spectrum-related issues have to be efficiently solved. This observation leads us to a third commandment, that is, innovative algorithms including exact and heuristic ones should be developed to tackle the various shades in applying network coding and network coding-related design problems. Otherwise, the potential gain enabled by network coding may not be fully realized. \\

\section{Re-designing Optical Transport Networks with Network Coding: Some Results}
In this part, we report some of our recent results regarding to the re-design of optical transport networks with network coding operations. We compare the network performances measured by routing cost, spectrum cost and network throughput. The network topology under test is a realistic one, COST239 as shown in Fig. 6 and Table \ref{tab: network} summarizes its key characteristics. All the results comparing network coding-enabled designs and conventional designs ones are based on solving optimally the network design formulation, often based on mixed integer linear programming models. Such comparison based on optimal values of both designs are to guarantee the fairness of benchmarking. Three technologies for optical core networks are covered including opaque WDM networks, transparent WDM ones and the recently proposed Elastic Optical Networks. 

\begin{figure}[!ht]
	\centering
	\includegraphics[width=0.7\linewidth, height = 5.5cm]{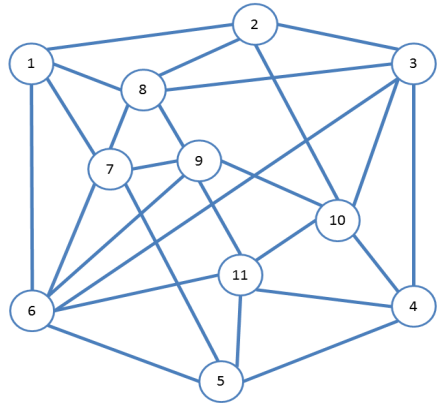}
	\caption{COST239 Network Topology}
	\label{fig:topology}
\end{figure}

\begin{table}[ht]
	\caption{Topology Characteristics}
	\label{tab: network}
	\centering
	\begin{tabular}{|c|c|}
		\hline
		\multicolumn{2}{|c|}{\small{Network Information}} \\
		\hline \hline
		\small{Nodes} & \small{11} \\ \hline
		\small{Links} & \small{26x2} \\ \hline
		\small{Node Degree 4} & \small{Node ID: 1, 2, 4, 5} \\ \hline
		\small{Node Degree 5} & \small{Node ID: 3, 7, 8, 9, 10, 11} \\ \hline
		\small{Node Degree 6} & \small{Node ID: 6} \\ \hline
	\end{tabular}
\end{table}

\subsection{Opaque WDM networks}
The opaque WDM networks have been remained widely deployed in optical core networks thanks to, on one hand, the historical factor, and on the other hand, its merit of regeneration capability at intermediate nodes. For accommodating traffic demands, the central issue to be solved is the routing problem and if network coding is enabled, a new problem arises, called, routing and network coding assignment problem (RNCA). In particular, in addition to the tradition task of identifying the optimal route for each demand, new opportunities and also challenges emerge with respect to determining pair of demands for encoding and decoding to reduce the overall traffic in the network. In \cite{hai_access, hai_springer2}, we provided the optimal design framework for solving the RNCA problem and Table \ref{tab: r1} highlights the comparison between the coding-aware designs and non-coding ones. It has been clearly seen that applying network coding brings about remarkable improvements in network performances including routing cost and network throughput. Besides, it has to be noted that such performance gain comes at a cost of increasing complexity for network designs. \\  

\begin{table}[!ht]
	\caption{A Comparison of Network Coding-enabled Design versus Traditional (Non-coding) Design in Opaque WDM Networks}
	\label{tab: r1}
	\centering
	\begin{tabular}{|c|c|}
		\hline
		\multicolumn{2}{|c|}{RNCA vs. Routing} \\
		\hline \hline
		Complexity (Variables) &  $O(|D||V||E|)$ vs. $O(|D||E|)$ \\ \hline
		Routing Cost & $\approx 30\%$ \cite{hai_access} \\ \hline
		Network Throughput & $\approx 30\%$ \cite{hai_springer2} \\ \hline
	\end{tabular}
\end{table}

\subsection {Transparent WDM networks}
Transparent WDM networks have long been progressing with the advances in transmission and switching technologies, allowing long-haul and ultra-high bit-rate operations. Optical core networks have then adopted this paradigm to partly replace the legacy opaque ones. Provisioning traffic demands in transparent WDM networks involves solving the routing and wavelength assignment (RWA) which is known as computationally hard. The arrival of network coding nevertheless adds a new layer of complexity for network designs, that is, the network coding assignment. In addition to the determination of coding nodes, pair of demands for encoding, the critical issue of wavelength assignment for a pair of demand before and after encoding must be taken into account. In \cite{hai_oft, hai_comletter}, an optimal framework for solving such issue (RWNCA) was proposed and Table \ref{tab: r2} draws a performance comparison between solving the RWNCA problem and the traditional RWA ones. It has been revealed that enabling coding functions in transparent WDM networks could bring about considerable gain in wavelength usage and network throughput. 

\begin{table}[!ht]
	\caption{A Comparison of Network Coding-enabled Design versus Traditional (Non-coding) Design in Transparent WDM Networks}
	\label{tab: r2}
	\centering
	\begin{tabular}{|c|c|}
		\hline
		\multicolumn{2}{|c|}{RWNCA vs. RWA} \\
		\hline \hline
		Complexity (Variables) &  $O(|D||V||E||W|)$ vs. $O(|D||E||W|)$ \\ \hline
		Wavelength Cost & $\approx 25\%$ \cite{hai_oft}  \\ \hline
		Network Throughput & $\approx 30\%$ \cite{hai_oft, hai_comletter} \\ \hline
	\end{tabular}
\end{table}

\subsection {Elastic Optical Networks}
Elastic optical networks (EONs) represent a future-proof framework as adaptive traffic provisioning depending on the specific bit-rate requirement and/or transmission quality of a demand is permitted. In doing so, they enables the use of fiber capacity more efficiently, therefore avoiding over-provision or under-utilizing the fiber spectrum. A central issue to be tackled in EONs is the routing and spectrum assignment (RSA) problem, that is, a more complicated version of RWA problem in WDM networks. In an analogy to the RWNCA problem, in network coding-enabled elastic optical networks, the routing, spectrum and network coding assignment (RSNCA) arises and have to be solved optimally to maximize the network coding benefits. Table \ref{tab: r3} compares the network throughput in traditional EONs and in network coding-enabled EONs and encouraging improvement up to roughly $30\%$ has been observed in our studies. 

\begin{table}[!ht]
	\caption{A Comparison of Network Coding-enabled Design versus Traditional (Non-coding) Design in Elastic Optical Networks}
	\label{tab: r3}
	\centering
	\begin{tabular}{|c|c|}
		\hline
		\multicolumn{2}{|c|}{RSNCA vs. RSA} \\
		\hline \hline
		Complexity (Variables) &  $O(|D||V||E||S|)$ vs. $O(|D||E||S|)$ \\ \hline
		Network Throughput &  $\approx 30\%$ \cite{hai_comcom2, hai_systems} \\ \hline
	\end{tabular}
\end{table}

\section{Summary}
This paper has discussed about the use of photonic network coding to transform the entire operations of existing optical networks infrastructure towards a greater capacity and security efficiency. This is enabled by upgrading the intermediate nodes functions with encoding/decoding capabilities in photonic domain. Three critical points have been raised addressing various shades of applying photonic NC including NC as a new dimension for multiplexing optical signals, NC as a radically efficient encryption tool at scale and algorithms aspects lie at the heart of realizing NC benefits.  \\

\section*{Acknowledgment}

This research is funded by the Vietnam National Foundation for Science and Technology Development (NAFOSTED) under the grant number 102.02-2018.09 for the project entitled, Network \textbf{C}oding for Spectrum-\textbf{E}fficient and Reliable \textbf{O}ptical Core Networks (CEO)

\bibliographystyle{IEEEtran}
\bibliography{IEEEabrv,ref}

\begin{thebibliography}{10}
\providecommand{\url}[1]{#1}
\csname url@samestyle\endcsname
\providecommand{\newblock}{\relax}
\providecommand{\bibinfo}[2]{#2}
\providecommand{\BIBentrySTDinterwordspacing}{\spaceskip=0pt\relax}
\providecommand{\BIBentryALTinterwordstretchfactor}{4}
\providecommand{\BIBentryALTinterwordspacing}{\spaceskip=\fontdimen2\font plus
\BIBentryALTinterwordstretchfactor\fontdimen3\font minus
  \fontdimen4\font\relax}
\providecommand{\BIBforeignlanguage}[2]{{%
\expandafter\ifx\csname l@#1\endcsname\relax
\typeout{** WARNING: IEEEtran.bst: No hyphenation pattern has been}%
\typeout{** loaded for the language `#1'. Using the pattern for}%
\typeout{** the default language instead.}%
\else
\language=\csname l@#1\endcsname
\fi
#2}}
\providecommand{\BIBdecl}{\relax}
\BIBdecl

\bibitem{ir4}
\BIBentryALTinterwordspacing
R.~Sabella, P.~Iovanna, G.~Bottari, and F.~Cavaliere, ``Optical transport for
  industry 4.0,'' \emph{J. Opt. Commun. Netw.}, vol.~12, no.~8, pp. 264--276,
  Aug 2020. [Online]. Available:
  \url{http://jocn.osa.org/abstract.cfm?URI=jocn-12-8-264}
\BIBentrySTDinterwordspacing

\bibitem{hai_oft2}
\BIBentryALTinterwordspacing
D.~T. Hai, ``On the spectrum-efficiency of qos-aware protection in elastic
  optical networks,'' \emph{Optik}, vol. 202, p. 163563, 2020. [Online].
  Available:
  \url{https://www.sciencedirect.com/science/article/pii/S0030402619314615}
\BIBentrySTDinterwordspacing

\bibitem{hai_optik}
\BIBentryALTinterwordspacing
D.~T. Hai, H.~T. Minh, and L.~H. Chau, ``Qos-aware protection in elastic
  optical networks with distance-adaptive and reconfigurable modulation
  formats,'' \emph{Optical Fiber Technology}, vol.~61, p. 102364, 2021.
  [Online]. Available:
  \url{https://www.sciencedirect.com/science/article/pii/S1068520020303540}
\BIBentrySTDinterwordspacing

\bibitem{hai_thesis}
\BIBentryALTinterwordspacing
H.~DAO~THANH, ``{Contribution to Flexible Optical Network Design: Spectrum
  Assignment and Protection},'' Theses, {T{\'e}l{\'e}com Bretagne ;
  Universit{\'e} de Bretagne Occidentale}, Mar. 2014. [Online]. Available:
  \url{https://hal.archives-ouvertes.fr/tel-01206788}
\BIBentrySTDinterwordspacing

\bibitem{hai_iet}
D.~T. {Hai}, M.~{Morvan}, and P.~{Gravey}, ``Combining heuristic and exact
  approaches for solving the routing and spectrum assignment problem,''
  \emph{IET Optoelectronics}, vol.~12, no.~2, pp. 65--72, 2018.

\bibitem{hai_wiley}
\BIBentryALTinterwordspacing
H.~Dao, M.~Morvan, and P.~Gravey, ``An efficient network-side path protection
  scheme in ofdm-based elastic optical networks,'' \emph{International Journal
  of Communication Systems}, vol.~31, no.~1, p. e3410, e3410 dac.3410.
  [Online]. Available:
  \url{https://onlinelibrary.wiley.com/doi/abs/10.1002/dac.3410}
\BIBentrySTDinterwordspacing

\bibitem{hai_ps1}
\BIBentryALTinterwordspacing
D.~Hai, M.~Morvan, and P.~Gravey, ``On the routing and spectrum assignment with
  multiple objectives,'' in \emph{Advanced Photonics for Communications}.\hskip
  1em plus 0.5em minus 0.4em\relax Optical Society of America, 2014, p.
  JT3A.12. [Online]. Available:
  \url{http://www.osapublishing.org/abstract.cfm?URI=PS-2014-JT3A.12}
\BIBentrySTDinterwordspacing

\bibitem{hai_ps2}
\BIBentryALTinterwordspacing
P.~Gravey, D.~Hai, and M.~Morvan, ``On the advantages of co-ofdm transponder in
  network-side protection,'' in \emph{Advanced Photonics for
  Communications}.\hskip 1em plus 0.5em minus 0.4em\relax Optical Society of
  America, 2014, p. PW1B.3. [Online]. Available:
  \url{http://www.osapublishing.org/abstract.cfm?URI=PS-2014-PW1B.3}
\BIBentrySTDinterwordspacing

\bibitem{hai_csndsp}
H.~{Dao Thanh}, M.~{Morvan}, and P.~{Gravey}, ``On the usage of flexible
  transponder in survivable transparent flex-grid optical network,'' in
  \emph{2014 9th International Symposium on Communication Systems, Networks
  Digital Sign (CSNDSP)}, July 2014, pp. 1123--1127.

\bibitem{hai_icist1}
D.~T. {Hai}, ``A novel adaptive operation of multi-line rate transponder for
  dedicated protection in wdm network,'' in \emph{2017 Seventh International
  Conference on Information Science and Technology (ICIST)}, April 2017, pp.
  69--74.

\bibitem{hai_icist2}
------, ``Multi-objective genetic algorithm for solving routing and spectrum
  assignment problem,'' in \emph{2017 Seventh International Conference on
  Information Science and Technology (ICIST)}, April 2017, pp. 177--180.

\bibitem{hai_sigtel1}
D.~T. {Hai} and K.~M. {Hoang}, ``An efficient genetic algorithm approach for
  solving routing and spectrum assignment problem,'' in \emph{2017
  International Conference on Recent Advances in Signal Processing,
  Telecommunications Computing (SigTelCom)}, Jan 2017, pp. 187--192.

\bibitem{hai_sigtel2}
------, ``On the efficient use of multi-line rate transponder for shared
  protection in wdm network,'' in \emph{2017 International Conference on Recent
  Advances in Signal Processing, Telecommunications Computing (SigTelCom)}, Jan
  2017, pp. 181--186.

\bibitem{hai_icact}
H.~D. {Thanh}, M.~{Morvan}, P.~{Gravey}, F.~{Cugini}, and I.~{Cerutti}, ``On
  the spectrum-efficiency of transparent optical transport network design with
  variable-rate forward error correction codes,'' in \emph{16th International
  Conference on Advanced Communication Technology}, Feb 2014, pp. 1173--1177.

\bibitem{hai_nics}
D.~M. Nguyen, L.~A. Ngoc, P.~T.~V. Huong, N.~H. Son, and D.~T. Hai, ``An
  efficient column generation approach for solving the routing and spectrum
  assignment problem in elastic optical networks,'' in \emph{2019 6th NAFOSTED
  Conference on Information and Computer Science (NICS)}, 2019, pp. 130--135.

\bibitem{hai2021development}
D.~T. Hai, L.~A. Ngoc, V.~N. Cham, and N.~Q. Cuong, ``On development of
  efficient data acquisition systems and parameter extraction technique for dfb
  lasers,'' 2021, arXiv:2105.06556.

\bibitem{20years}
\BIBentryALTinterwordspacing
P.~J. Winzer, D.~T. Neilson, and A.~R. Chraplyvy, ``Fiber-optic transmission
  and networking: the previous 20 and the next 20 years,'' \emph{Opt. Express},
  vol.~26, no.~18, pp. 24\,190--24\,239, Sep 2018. [Online]. Available:
  \url{http://www.opticsexpress.org/abstract.cfm?URI=oe-26-18-24190}
\BIBentrySTDinterwordspacing

\bibitem{xor3}
L.~K. Chen, M.~Li, and S.~C. Liew, ``Breakthroughs in photonics 2014: Optical
  physical-layer network coding, recent developments, and challenges,''
  \emph{IEEE Photonics Journal}, vol.~7, no.~3, pp. 1--6, June 2015.

\bibitem{crunch}
\BIBentryALTinterwordspacing
A.~D. Ellis, N.~M. Suibhne, D.~Saad, and D.~N. Payne, ``Communication networks
  beyond the capacity crunch,'' \emph{Philosophical Transactions of the Royal
  Society of London A: Mathematical, Physical and Engineering Sciences}, vol.
  374, no. 2062, 2016. [Online]. Available:
  \url{http://rsta.royalsocietypublishing.org/content/374/2062/20150191}
\BIBentrySTDinterwordspacing

\bibitem{NC}
R.~Ahlswede \emph{et~al.}, ``Network information flow,'' \emph{Information
  Theory, IEEE Transactions on}, vol.~46, no.~4, pp. 1204--1216, Jul 2000.

\bibitem{hai_comletter}
D.~T. Hai, ``Leveraging the survivable all-optical wdm network design with
  network coding assignment,'' \emph{IEEE Communications Letters}, vol.~21,
  no.~10, pp. 2190--2193, Oct 2017.

\bibitem{hai_access}
------, ``An optimal design framework for 1+1 routing and network coding
  assignment problem in wdm optical networks,'' \emph{IEEE Access}, vol.~5, pp.
  22\,291--22\,298, 2017.

\bibitem{hai_oft}
\BIBentryALTinterwordspacing
T.~H. Dao, ``On optimal designs of transparent wdm networks with 1+1 protection
  leveraged by all-optical xor network coding schemes,'' \emph{Optical Fiber
  Technology}, vol.~40, pp. 93 -- 100, 2018. [Online]. Available:
  \url{http://www.sciencedirect.com/science/article/pii/S1068520017303012}
\BIBentrySTDinterwordspacing

\bibitem{hai_comcom}
\BIBentryALTinterwordspacing
D.~T. Hai, ``A bi-objective integer linear programming model for the routing
  and network coding assignment problem in wdm optical networks with dedicated
  protection,'' \emph{Computer Communications}, vol. 133, pp. 51 -- 58, 2019.
  [Online]. Available:
  \url{http://www.sciencedirect.com/science/article/pii/S0140366418300148}
\BIBentrySTDinterwordspacing

\bibitem{hai_comcom2}
\BIBentryALTinterwordspacing
------, ``On routing, spectrum and network coding assignment problem for
  transparent flex-grid optical networks with dedicated protection,''
  \emph{Computer Communications}, 2019. [Online]. Available:
  \url{http://www.sciencedirect.com/science/article/pii/S0140366418306546}
\BIBentrySTDinterwordspacing

\bibitem{hai_springer}
\BIBentryALTinterwordspacing
------, ``On solving the 1 + 1 routing, wavelength and network coding
  assignment problem with a bi-objective integer linear programming model,''
  \emph{Telecommunication Systems}, vol.~71, no.~2, pp. 155--165, Jun 2019.
  [Online]. Available: \url{https://doi.org/10.1007/s11235-018-0474-9}
\BIBentrySTDinterwordspacing

\bibitem{hai_rtuwo}
D.~T. {Hai}, ``Re-designing dedicated protection in transparent wdm optical
  networks with xor network coding,'' in \emph{2018 Advances in Wireless and
  Optical Communications (RTUWO)}, Nov 2018, pp. 118--123.

\bibitem{hai_systems}
D.~T. Hai, L.~H. Chau, and N.~T. Hung, ``A priority-based multiobjective design
  for routing, spectrum, and network coding assignment problem in
  network-coding-enabled elastic optical networks,'' \emph{IEEE Systems
  Journal}, vol.~14, no.~2, pp. 2358--2369, 2020.

\bibitem{hai_springer2}
\BIBentryALTinterwordspacing
D.~T. Hai, ``Network coding for improving throughput in wdm optical networks
  with dedicated protection,'' \emph{Optical and Quantum Electronics}, vol.~51,
  no. 387, November 2019. [Online]. Available:
  \url{https://doi.org/10.1007/s11082-019-2104-5}
\BIBentrySTDinterwordspacing

\bibitem{hai2021consolation}
------, ``The consolation of network coding and partial protection techniques
  to optical transport networks in data, data, data era,'' 2021,
  arXiv:2105.03503.

\bibitem{nc_others9}
\BIBentryALTinterwordspacing
S.~Li, E.~W.~M. Wong, H.~{\O}verby, and M.~Zukerman, ``Performance modeling of
  diversity coded path protection in obs/ops networks,'' \emph{J. Lightwave
  Technol.}, vol.~37, no.~13, pp. 3138--3152, Jul 2019. [Online]. Available:
  \url{http://jlt.osa.org/abstract.cfm?URI=jlt-37-13-3138}
\BIBentrySTDinterwordspacing

\bibitem{nc_others7}
Y.~Su, X.~Meng, Q.~Kang, and X.~Han, ``Survivable virtual network link
  protection method based on network coding and protection circuit,''
  \emph{IEEE Access}, vol.~6, pp. 67\,477--67\,493, 2018.

\bibitem{icc}
H.~Overby \emph{et~al.}, ``Cost comparison of 1+1 path protection schemes: A
  case for coding,'' in \emph{ICC 2012, IEEE}, June 2012, pp. 3067--3072.

\bibitem{nc_others1}
\BIBentryALTinterwordspacing
C.~Mitsolidou, N.~Pleros, and A.~Miliou, ``Digital all-optical physical-layer
  network coding for 2gbaud dqpsk signals in mm-wave radio-over-fiber
  networks,'' \emph{Optical Switching and Networking}, vol.~33, pp. 199--207,
  2019. [Online]. Available:
  \url{https://www.sciencedirect.com/science/article/pii/S1573427717300875}
\BIBentrySTDinterwordspacing

\bibitem{nc_others2}
X.~Guan, Q.~Yang, T.~Wang, and C.~C.-K. Chan, ``Phase-aligned physical-layer
  network coding in visible light communications,'' \emph{IEEE Photonics
  Journal}, vol.~11, no.~2, pp. 1--9, 2019.

\bibitem{nc_others3}
J.~Yanmei, L.~Congmin, S.~Pengfei, and L.~Lu, ``Modulated retro-reflector-based
  physical-layer network coding for space optical communications,'' \emph{IEEE
  Access}, vol.~9, pp. 44\,868--44\,880, 2021.

\bibitem{nc_others4}
\BIBentryALTinterwordspacing
N.~Feng and X.~Sun, ``Implementation of network-coding approach for improving
  the ber performance in non-orthogonal multiple access (noma)-pon,''
  \emph{Optics Communications}, vol. 462, p. 125301, 2020. [Online]. Available:
  \url{https://www.sciencedirect.com/science/article/pii/S0030401820300365}
\BIBentrySTDinterwordspacing

\bibitem{nc_others5}
\BIBentryALTinterwordspacing
R.~Lin, Y.~Cheng, X.~Guan, M.~Tang, D.~Liu, C.-K. Chan, and J.~Chen,
  ``Physical-layer network coding for passive optical interconnect in
  datacenter networks,'' \emph{Opt. Express}, vol.~25, no.~15, pp.
  17\,788--17\,797, Jul 2017. [Online]. Available:
  \url{http://www.opticsexpress.org/abstract.cfm?URI=oe-25-15-17788}
\BIBentrySTDinterwordspacing

\bibitem{nc_others6}
A.~Engelmann, W.~Bziuk, A.~Jukan, and M.~Médard, ``Exploiting parallelism with
  random linear network coding in high-speed ethernet systems,'' \emph{IEEE/ACM
  Transactions on Networking}, vol.~26, no.~6, pp. 2829--2842, 2018.

\bibitem{nc_others8}
N.~B. El~Asghar, I.~Jouili, and M.~Frikha, ``Survivable inter-datacenter
  network design based on network coding,'' in \emph{2017 IEEE/ACS 14th
  International Conference on Computer Systems and Applications (AICCSA)},
  2017, pp. 1192--1197.

\bibitem{nc_security1}
\BIBentryALTinterwordspacing
G.~Savva, K.~Manousakis, and G.~Ellinas, ``Confidentiality meets protection in
  elastic optical networks,'' \emph{Optical Switching and Networking}, p.
  100620, 2021. [Online]. Available:
  \url{https://www.sciencedirect.com/science/article/pii/S1573427721000175}
\BIBentrySTDinterwordspacing

\bibitem{nc_security2}
------, ``Providing confidentiality in optical networks: Metaheuristic
  techniques for the joint network coding-routing and spectrum allocation
  problem,'' in \emph{2020 22nd International Conference on Transparent Optical
  Networks (ICTON)}, 2020, pp. 1--4.

\bibitem{nc_others10}
A.~Kotb, K.~E. Zoiros, and C.~Guo, ``1 tb/s all-optical xor and and gates using
  quantum-dot semiconductor optical amplifier-based turbo-switched
  mach–zehnder interferometer,'' \emph{Journal of Computational Electronics},
  2019.

\bibitem{nc_others11}
K.~MISHINA, D.~HISANO, and A.~MARUTA, ``All-optical modulation format
  conversion and applications in future photonic networks,'' \emph{IEICE
  Transactions on Electronics}, vol. E102.C, no.~4, pp. 304--315, 2019.

\bibitem{hai2021shades}
D.~T. Hai, ``Three shades of partial protection in elastic optical networks,''
  2021, arXiv:2105.01046.

\bibitem{hai2021achilles}
------, ``The achilles heel of some optical network designs and performance
  comparisons,'' 2021, arXiv:2105.07088.

\bibitem{hai_soict}
\BIBentryALTinterwordspacing
H.-C. Le, H.~Dao-Thanh, and N.~T. Dang, ``Development of dynamic qot-aware
  lightpath provisioning scheme with flexible advanced reservation for
  distributed multi-domain elastic optical networks,'' in \emph{Proceedings of
  the Tenth International Symposium on Information and Communication
  Technology}, ser. SoICT 2019.\hskip 1em plus 0.5em minus 0.4em\relax New
  York, NY, USA: Association for Computing Machinery, 2019, p. 210–215.
  [Online]. Available: \url{https://doi.org/10.1145/3368926.3369723}
\BIBentrySTDinterwordspacing

\bibitem{hai2021pragmatic}
D.~T. Hai and L.~A. Ngoc, ``A pragmatic approach for designing transparent wdm
  optical networks with multi-objectives,'' 2021, arXiv:2105.08588.

\end{thebibliography}

\vspace{12pt}
\color{red}

\end{document}